\documentclass[sigconf]{acmart}

\usepackage{listings}
\usepackage{xcolor}

\definecolor{codegreen}{rgb}{0,0.6,0}
\definecolor{codegray}{rgb}{0.5,0.5,0.5}
\definecolor{codepurple}{rgb}{0.58,0,0.82}
\definecolor{backcolour}{rgb}{0.95,0.95,0.92}

\lstdefinestyle{mystyle}{
backgroundcolor=\color{backcolour},
commentstyle=\color{codegreen},
keywordstyle=\color{magenta},
numberstyle=\tiny\color{codegray},
stringstyle=\color{codepurple},
basicstyle=\ttfamily\footnotesize,
breakatwhitespace=false,
breaklines=true,
captionpos=b,
keepspaces=true,
numbers=left,
numbersep=5pt,
showspaces=false,
showstringspaces=false,
showtabs=false,
tabsize=2
}
\setcopyright{none} 
\renewcommand\footnotetextcopyrightpermission[1]{}
\begin{document}

\title{Shai-am: A Machine Learning Platform for Investment Strategies}

\author{Jonghun Kwak}
\orcid{0000-0003-4291-5033}
\affiliation{%
\institution{Shinhan AI}
\streetaddress{Yeui-daero 70}
\city{Seoul}
\country{Republic of Korea}
\postcode{07325}
}
\email{jkwak3@berkeley.edu}

\author{Jungyu Ahn}
\affiliation{%
\institution{Shinhan AI}
\streetaddress{Yeui-daero 70}
\city{Seoul}
\country{Republic of Korea}
\postcode{07325}
}
\email{jungyuahn@shinhan.com}

\author{Jinho Lee}
\affiliation{%
\institution{Shinhan AI}
\streetaddress{Yeui-daero 70}
\city{Seoul}
\country{Republic of Korea}
\postcode{07325}
}
\email{jinholee@shinhan.com}

\author{Sungwoo Park}
\affiliation{%
\institution{Shinhan AI}
\streetaddress{Yeui-daero 70}
\city{Seoul}
\country{Republic of Korea}
\postcode{07325}
}
\email{sungwoopark@shinhan.com}

\renewcommand{\shortauthors}{Jonghun Kwak, et al.}

\begin{abstract}
The finance industry has adopted machine learning (ML) as a form of quantitative research to support better investment decisions, yet there are several challenges often overlooked in practice. (1) ML code tends to be unstructured and ad hoc, which hinders cooperation with others. (2) Resource requirements and dependencies vary depending on which algorithm is used, so a flexible and scalable system is needed. (3) It is difficult for domain experts in traditional finance to apply their experience and knowledge in ML-based strategies unless they acquire expertise in recent technologies. This paper presents Shai-am, an ML platform integrated with our own Python framework. The platform leverages existing modern open-source technologies, managing containerized pipelines for ML-based strategies with unified interfaces to solve the aforementioned issues. Each strategy implements the interface defined in the core framework. The framework is designed to enhance reusability and readability, facilitating collaborative work in quantitative research. Shai-am aims to be a pure AI asset manager for solving various tasks in financial markets.
\end{abstract}


\maketitle
\pagestyle{plain}

\section{Introduction}
The application of machine learning (ML) to solve complex problems is now pervasive across domains, and the finance industry is not an exception. As Krishnamachari \cite{krishnamachari2017big} noted, the role of humans in the finance industry has already declined significantly since the big data and AI revolution. Many prior works have attempted to apply various ML algorithms in the finance domain. Kim and Won \cite{kim2018forecasting} proposed long short-term memory (LSTM) with multiple generalized autoregressive conditional heteroskedasticity (GARCH) models to predict the volatility of the stock price index. Lei \textit{et al.} \cite{lei2020time} used both supervised learning and reinforcement learning to solve financial signal representation and algorithmic trading problems.

However, most studies have focused on the methodology itself, not on how to actually deploy models in real-world systems. As Sculley \textit{et al.} \cite{sculley2015hidden} asserted, it is important to be aware of ML-specific risk factors in systems, such as boundary erosion, data dependencies, and a variety of anti-patterns. They encouraged better abstractions, testing methodologies, and design patterns in ML systems to avoid the incurrence of massive maintenance costs. To avoid the possible issues, many ML platforms, such as Bighead \cite{brumbaugh2019bighead} and NSML \cite{sung2017nsml}, have been presented by various organizations and individuals. However, these platforms are limited to developing and deploying ML models for general purposes. Few platforms have aimed to unlock the power of ML in quantitative investment, as Qlib \cite{yang2020qlib} did.

Thus, we propose an ML platform called Shai-am for developing and deploying ML-based investment strategies. It aims to abstract strategic logic into isolated components to maximize reusability. We design the core framework in Python for implementing investment strategies to expose unified interfaces. Moreover, we aim to allow ML or quantitative researchers to focus solely on core logic without deeply concerning themselves with resources and dependencies. Shai-am leverages various open-source projects to enable such an environment; designing pipelines and allocating resources can be done flexibly in the platform. This paper explores Shai-am's overall design goals and architecture and how it can be utilized in financial business.

\section{Background}
In this section, we review the open-source projects that we use in our platform and other related works.
\subsection{Kubernetes}
Kubernetes \cite{bernstein2014containers} is an open-source container orchestration system for deploying and managing containerized applications. It abstracts the underlying infrastructure, managing and scaling resources automatically. Recently, many previous works have taken advantage of it to build ML or data analysis systems. For example, Lee \textit{et al.} \cite{lee2020multi} proposed a cloud native ML platform with dynamic orchestration and microservice-oriented containerization using Kubernetes. They improved the efficiency of computing resources and solved the difficulty of maintaining ML development environments. Tesliuk \textit{et al.} \cite{tesliuk2019kubernetes} developed a data analysis pipeline on Kubernetes for single particle imaging (SPI) experiments. Flexible modification or replacement of pipeline components and scaling of individual components was possible due to the nature of Kubernetes. Katib \cite{tesliuk2019kubernetes}, a scalable automatic machine learning (AutoML) platform, is also built on Kubernetes, encapsulating each component as a microservice.

\subsection{Airflow}
Apache Airflow \cite{airflow} is an open-source workflow platform known for its scalability and extensibility. It aims to be a Kubernetes-friendly project to take advantage of Kubernetes' increased stability and autoscaling abilities. In addition, Airflow allows users to define and schedule jobs programmatically in Python, allowing for dynamic pipeline generation \cite{airflow}. Although Airflow is a general solution for controlling workflows, it is often used to construct ML pipelines. Bighead \cite{brumbaugh2019bighead}, a framework-agnotic, end-to-end ML platform developed by Airbnb, exploited the virtues of Airflow. Its offline execution engine is built on top of Airflow, managing training, inference, and evaluation jobs.

\subsection{MinIO}
MinIO \cite{minio} is a high-performance object storage that is compatible with Amazon S3 cloud storage \cite{s3}. It is also Kubernetes native, offering consistent performance in diverse environments. MinIO can store unstructured data, such as images, videos, log files, and even trained ML models. Thus, several prior works have used MinIO in their ML experiments. For instance, Buniatyan \cite{buniatyan2019hyper}, proposing a methodology for processing input data of deep learning models, stored his training data and artifacts in MinIO. Baxevanakis \textit{et al.} \cite{baxevanakis2022mever} built a DeepFake detection service based on a deep learning model using MinIO for its main storage.

\subsection{Related Work}
\subsubsection{API Design}
In general, application programming interface (API) design is essential because it is closely related to the usability of the API. A well-designed API enables users to understand the whole architecture easily and reduces the overall complexity. It also facilitates the development and maintenance of any kind of project. One example is Scikit-learn \cite{pedregosa2011scikit}, an open-source Python library for solving medium-scale supervised and unsupervised ML problems. It provides simple and efficient interfaces so that it can be integrated easily into applications in various domains \cite{buitinck2013api}. Sktime \cite{loning2019sktime} is a Scikit-learn compatible Python library that focuses on ML with time series. L{\"o}ning \textit{et al.} \cite{loning2019sktime} emphasized the importance of a unified interface and the way that their design impacts the solution of time series tasks.

\subsubsection{ML Platforms for Finance}
Many organizations and individuals have built ML platforms to reduce the pain of the whole ML modeling process. Airbnb introduced Bighead \cite{brumbaugh2019bighead} to help its data science and engineering teams develop and deploy ML models efficiently. Similarly, NSML \cite{sung2017nsml} was made to improve the overall efficiency of model development. Most of the works are oriented to the general ML development process, and few platforms have been built to deal with ML in finance specifically. Qlib \cite{yang2020qlib} is one of few ML platforms focusing on quantitative investment. Its goal is to empower quantitative researchers to take full advantage of ML technologies in investment.

\subsubsection{Backtesting}
Backtesting refers to a methodology for testing an investment strategy based on historical data. There are several open-source backtesting modules built in Python, such as Backtrader \cite{backtrader} and Zipline \cite{zipline}. Previous studies have utilized these modules to evaluate their models. Fazeli \cite{fazeli2019using} used deep learning for predicting stock trends and backtested it using Backtrader. Picasso \textit{et al.} \cite{picasso2019technical} also used Backtrader to evaluate the market prediction of their deep learning model. Joshi \cite{joshi2018developing} proposed an LSTM to predict the next day's stock price and verified his idea with Zipline.

\section{Overall System Design}
\subsection{Infrastructure and Modeling Environment}
We set up a Kubernetes cluster \cite{bernstein2014containers} on premise, installing a private Docker registry, Apache Airflow \cite{airflow}, and MinIO \cite{minio}. The Docker registry stores the images for executing strategy instances. We choose Airflow as our workflow platform to schedule and run the containerized pipelines for ML-based strategies. Before deploying actual pipelines, analyzing data and conducting experiments are mandatory to develop an investment strategy. For such an environment, we launch a Kubernetes service with the same image used in the pipeline for each researcher. A researcher accesses the container on the cluster remotely, performs various experiments, and pushes final implementations to the remote Git repository. The object storage, MinIO, is for saving logs of pipelines, artifacts and, most importantly, strategy instances. The overall modeling environment in Shai-am is shown in Figure 1.

\begin{figure}[h]
  \centering
  \includegraphics[width=\linewidth]{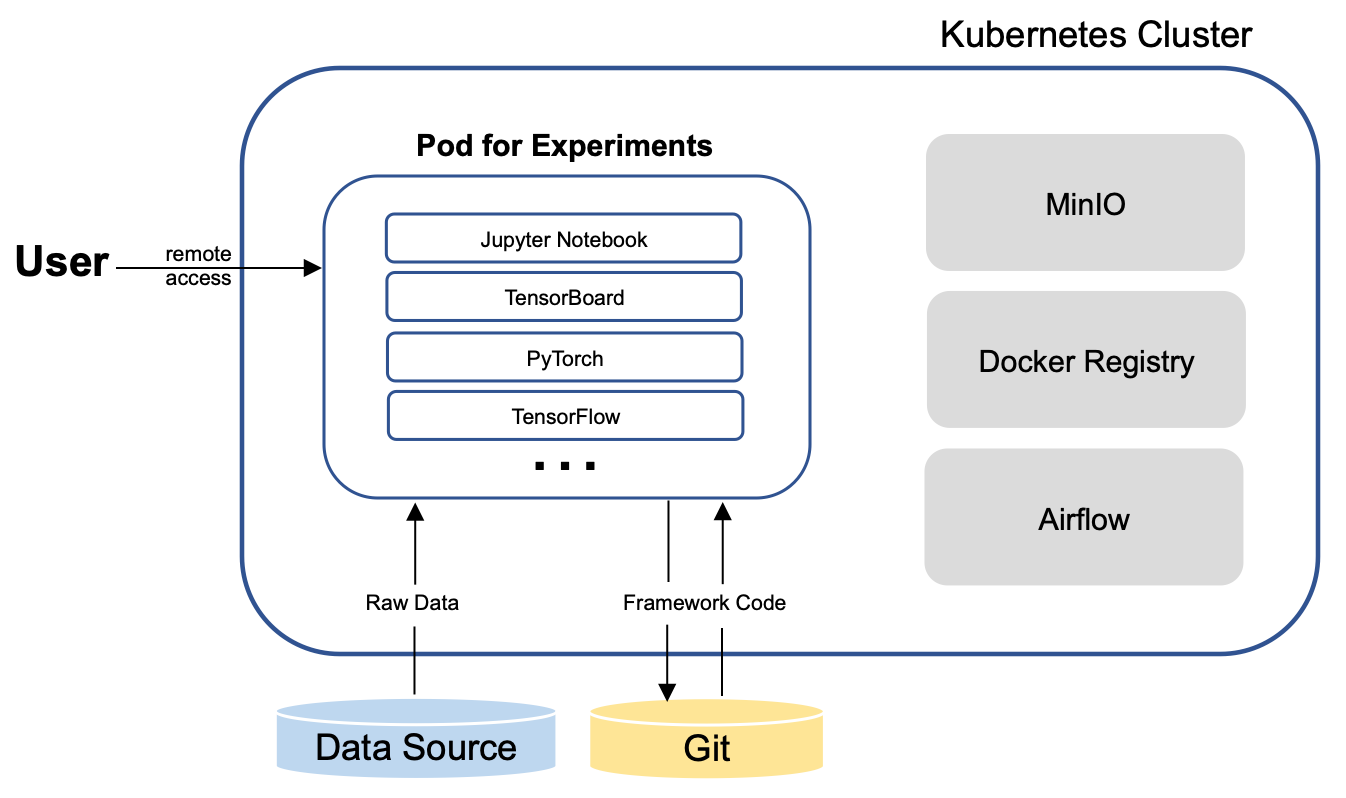}
\caption{Modeling environment}
\end{figure}

\subsection{Framework for Investment Strategies}
Python is the most preferred language for the fields of machine learning and data analysis, offering clean high-level APIs \cite{raschka2020machine}. To accommodate other popular ML frameworks or libraries in Python, such as TensorFlow \cite{abadi2016tensorflow}, PyTorch \cite{paszke2017automatic}, and Scikit-learn \cite{pedregosa2011scikit}, we choose it as the main language. In addition, our primary focus is not high frequency trading (HFT) or intraday trading; run-time optimization is not a critical issue. Our ML-based strategies aim to replace the decision-making process of traditional asset managers in investment. Our core interfaces are designed to maximize reusability and minimize the silo effect. Further details of the interfaces are elaborated in Section 4. All source codes, including the interfaces and implementations of our investment strategies, are managed with Git as the complete Python project. One can easily clone the repository and install it as a Python package.

\begin{figure*}[ht]
  \centering
  \includegraphics[width=\textwidth]{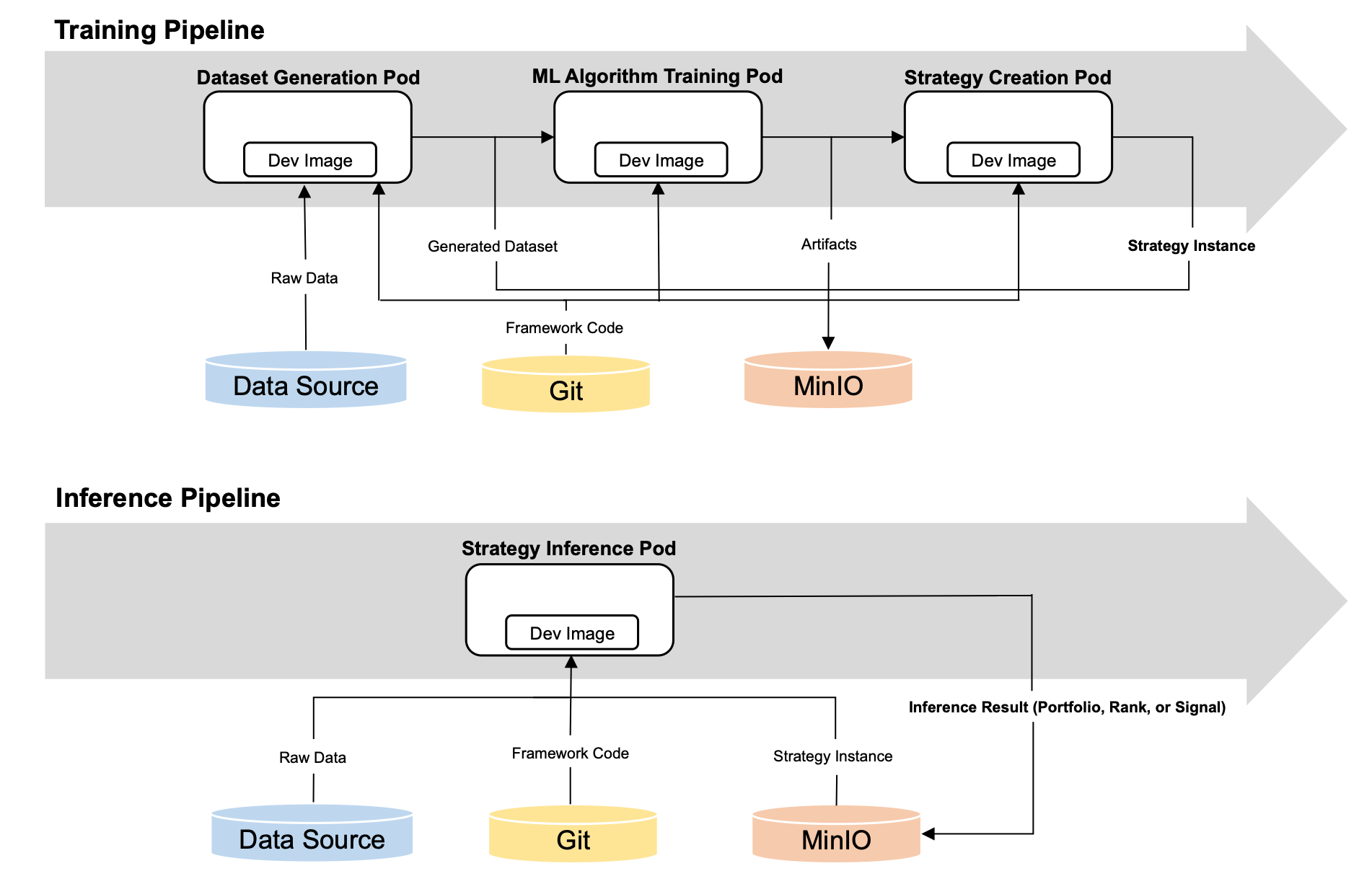}
\caption{Pipelines for an investment product}
\end{figure*}

\subsection{ML Pipeline}
Our ML pipeline is defined as Airflow's directed acyclic graph (DAG). We configure Airflow with \textit{KubernetesExecutor} and define each task using \textit{KubernetesPodOperator}. The task runs a Python script importing our core framework package. Since each component of the pipeline is executed as a Kubernetes pod, configuring the execution environment, including Docker image and resources (cpu, gpu, memory, etc.), is flexible. After a train pipeline runs successfully, a trained ML-based strategy instance is saved in MinIO with a unique identification number. Due to the volatile nature of the pod, anything needed to be archived, such as a log, artifact, or strategy, is saved to MinIO. A pipeline for inference is scheduled appropriately depending on the properties of the strategy. When the pipeline runs, it loads the strategy object from the storage and executes it as described in Section 4.3. The output of the strategy, such as a portfolio, is also saved in MinIO.

\section{Core Framework Design}
\subsection{Dataset}
We unify the process of loading data from the remote data source by developing the loader module. By calling the module, researchers can load data easily in Python without knowing the detailed structure of the data source. Based on the loader module, one needs to implement the \textit{BaseDataPipeline} interface to generate any data needed for an investment strategy. \textit{BaseTimeSeriesDatasetGenerator} extends the base interface and is oriented toward generating a time series dataset. The generated dataset must be a \textit{TimeSeriesDataset} object, which is likely to be an input of the algorithm for the time series. Since the type of time series dataset is identical, time series--related operations, such as splitting the dataset based on the date, can be performed by calling the common methods. A unified interface for generating a dataset also helps others understand how data are preprocessed and infer the characteristics of the algorithms that will use the dataset.

\subsection{Algorithm}
The objective of algorithm interfaces is to encapsulate the core logic and distinguish a broad category of algorithms while leaving maximum flexibility to researchers. Quantitative investment logic can be any type of algorithm; however, we want to identify its characteristics by defining a general interface for each kind. The \textit{BaseAlgorithm} interface is at the core of the algorithm interfaces. It requires the implementation of save and load methods for saving and loading an object. The \textit{BaseRLAlgorithm} interface extends to expose methods related to reinforcement learning. Moreover, an environment for the reinforcement learning algorithm must implement the interface defined in OpenAI Gym \cite{brockman2016openai}. Another child interface, \textit{BaseEstimator}, is for supervised and unsupervised learning algorithms, similar to the \textit{estimator} interface introduced in Scikit-learn. We also apply its key concepts of using mix-in classes and multi-inheritance \cite{buitinck2013api}.

\subsection{Strategy}
The \textit{BaseStrategy} interface is the most essential part of our investment framework. The interface is designed to be as simple as possible so that even a person without great expertise in Python can carry out a strategy. We expect domain experts in traditional finance, such as traders, managers, and even research analysts, to contribute to implementing strategies based on the framework. In addition to methods for saving and loading an object, the interface exposes two main methods: \textit{reset} and \textit{execute}. A \textit{reset} method is for setting the date range for which the strategy is to be executed. After the \textit{reset} is called, valid dates on which the strategy can be carried out are returned. Any date in the list of the valid dates is passed to call execute. The output of \textit{execute} must be an outcome instance that has content and extra information such as an investment horizon. A type of content varies depending on the type of strategy. For example, for an allocation strategy, a content of its outcome must be a portfolio instance that has information about assets and their weights. The interface also requires passing an investment universe, benchmark, and type of strategy to a constructor. These attributes are useful at the conceptual level to identify a strategy's basic characteristics. After instantiation, such information can be retrieved easily by calling the getters.

The \textit{BaseDPStrategy} interface extends the base interface to represent a strategy that utilizes data from the remote data source. A list of \textit{BaseDataPipeline} objects that are needed to execute a strategy must be passed to the constructor and can be accessed by calling the getter after instantiation. \textit{BaseDPStrategy} adds the \textit{configs} setter to configure the \textit{BaseDataPipeline} objects. Finally, the \textit{BaseMLStrategy} interface extends the \textit{BaseDPStrategy} to embrace ML algorithms for an investment strategy. A list of \textit{BaseAlgorithm} objects is required for the constructor. A \textit{BaseDataPipeline} object generates data for an algorithm; however, there is no direct dependency between two objects in terms of implementation. The strategy interface is the highest-level interface encapsulating the core logic for investment and the way in which the required data are generated. Executing the saved strategy object is straightforward, as shown in Listing 1.

\begin{lstlisting}[language=Python, caption=Code snippet for executing a strategy]
from shai_am.strategy.allocation import MyAllocStrategy
from shai_am.data.db.config import MY_CONFIG
from datetime import date

s = MyAllocStrategy.load(SAVED_PATH)
s.configs = MY_CONFIG #config for the remote data source
s.reset(date(2022, 1, 1), date.today())
outcome = s.execute(date.today())
today_port = outcome.content

\end{lstlisting}

\subsection{Backtesting}
The backtesting module exploits the fact that all strategies implement the same interface described in Section 4.3. This module unifies the process of backtesting an investment strategy, mediating the burden of writing custom codes for each one. The key aspect of the module is that its implementation is totally separated from how the strategy is implemented. It only calls its getters and methods defined in the strategy interface. Based on the strategy's investment universe and benchmark, the module queries price data and metadata from the remote data source. The loaded data are used to evaluate the strategy both horizontally and vertically. A horizontal condition determines whether the strategy is analyzed at monthly, quarterly, or yearly frequency, or in a customized time interval. In addition, the module is able to break down the performance of the strategy into individual assets, sectors, broad categories, or countries, depending on the given vertical condition.

\section{Use Case}
Shai-am is adopted as the main platform in managing ML-based investment products, such as open-ended funds, trusts, and wrap accounts. For each product, at least two pipelines are deployed, as shown in Figure 2: one for creating a strategy instance corresponding to the product and the other for actually carrying out the strategy. All selected strategies and their outcomes, such as portfolios, ranks, and signals, are saved and managed in MinIO. Saved results are used to make decisions on portfolio rebalancing, asset selection, or hedging. The backtesting module periodically loads the saved objects and evaluates their recent performances in diverse aspects. Researchers monitor the backtesting results on a daily basis to see if strategies are working as intended. In this way, human intervention is minimized, and the number of investment products being managed is no longer a critical issue.

\section{Conclusion}
In this paper, we present Shai-am, a scalable ML platform specialized in managing ML-based investment strategies. It is designed to maximize the productivity of researchers and developers while minimizing the risk of technical debt. We focus on abstracting quantitative investment logic into functionally isolated modules to increase reusability and foster a collaborative environment. We will continue working on Shai-am to seek better investment strategies based on modern ML technologies.


\end{document}